\documentclass[fleqn,twoside]{article}
\usepackage{espcrc2}
\usepackage{epsfig}
\readRCS
$Id: espcrc2.tex,v 1.2 2004/02/24 11:22:11 spepping Exp $
\newcommand{\AmS}{{\protect\the\textfont2
  A\kern-.1667em\lower.5ex\hbox{M}\kern-.125emS}}

\newcommand{\be}{\begin{equation}}
\newcommand{\ee}{\end{equation}}
\newcommand{\bea}{\begin{eqnarray}}
\newcommand{\eea}{\end{eqnarray}}

\title{A numerical approach to infrared divergent multi-parton \\
phase space integrals}

\author{G.~Heinrich\address{
II. Institut f\"ur Theoretische Physik, Universit\"at Hamburg,\\
Luruper Chaussee 149, 22761 Hamburg, Germany}
        \thanks{To appear in the proceedings of the Conference "Loops and 
	Legs in Quantum Field Theory", Zinnowitz, Germany, April 2004.}	}

\begin{document}

\begin{abstract}
It is described how the method of sector decomposition can serve to
disentangle overlapping infrared singularities, in particular those 
occurring in the calculation of the real emission part 
of $e^+e^- \to 2$\,jets and $e^+e^- \to 3$\,jets at NNLO.
\end{abstract}

\maketitle

\section{INTRODUCTION}

Particle physics nowadays has largely become a matter of high precision
measurements, which on the theory side requires an increasing number of 
loops and legs to be included in the calculations. 

The process  $e^+e^- \to $\,jets is particularly interesting, 
from an experimental as well as a theoretical point of view, 
because of its "clean" initial state, 
such that it can serve for a very accurate determination of $\alpha_s$.
However, LEP experiments already have shown that theoretical predictions at
next-to-leading order (NLO) are not always sufficient to match the experimental
precision\,\cite{bethke}, and this of course will be even more true for a 
future Linear Collider.  

These facts have triggered a lot of progress  in the calculation 
of NNLO corrections in recent years\,\cite{nigel}. The last missing 
piece for the construction of an NNLO Monte Carlo program 
for the process $e^+e^- \to 3$\,jets is in fact the double real 
radiation part, which involves phase space integrations over five partons, 
where up to two of them can become unresolved, 
leading to infrared singularities. 

The conventional way to deal with these singularities is to establish a
subtraction scheme to isolate the divergent
part\,\cite{Giele:1991vf,Frixione:1995ms,Catani:1996vz}. 
The latter then is calculated analytically in $D=4-2\epsilon$ dimensions,
leading to $1/\epsilon$ poles which 
will cancel against the ones from the virtual corrections. 
This procedure has been applied very successfully in NLO calculations. 
Its generalization to NNLO however is far from being straightforward. 
Nevertheless, subtraction schemes have been proposed in the 
literature
\cite{Kosower:2002su,wz2,Gehrmann-DeRidder:2004tv,Kilgore:2004ty,weinzierlll04}, 
whereas  
the problem of integrating the subtraction terms analytically
in $D$ dimensions has been solved only for the case  
$e^+e^- \to 2$\,jets so 
far\,\cite{Weinzierl:2003ra,Gehrmann-DeRidder:2003bm,Gehrmann-DeRidder:2004tv}.

What will be suggested here is a new method which does not rely on 
explicit subtraction terms. The infrared singularities are isolated 
in an automated way using sector 
decomposition\,\cite{secdec,Binoth:2000ps}.
The cancellation of the pole coefficients  with the ones from the 
virtual corrections can be verified numerically. 
The method already has been applied successfully to the process 
$e^+e^-\to 2$\,jets\,
\cite{Gehrmann-DeRidder:2003bm,Anastasiou:2003gr,Binoth:2004jv,Anastasiou:2004qd}.

\section{SECTOR DECOMPOSITION}

The method of sector decomposition acts on parameter integrals 
and serves to factorize singularities which have an overlapping 
structure, as in the following simple example:\\
$I=\int_0^1 dx_1\,dx_2 \,x_1^{-1-\epsilon}\,[x_1+x_2]^{-1}\;.$
Decomposing the parameter space into two sectors where the integration 
variables are ordered and remapping the integration range to the unit 
square factorizes the singularity:
$$I=\int_0^1 \frac{dx_1\,dx_2}{x_1+x_2} \,x_1^{-1-\epsilon}
\,[\underbrace{\Theta(x_1-x_2)}_{(1)}+\underbrace{\Theta(x_2-x_1)}_{(2)}]
$$
The substitution $x_2=x_1\,t_2$ in sector (1) and $x_1=x_2\,t_1$ 
in sector (2) leads to
\begin{eqnarray*}
I&=&\int_0^1 dx_1\,x_1^{-1-\epsilon}\int_0^1 dt_2
\,(1+t_2)^{-1}\\
&+&\int_0^1 dx_2
\,x_2^{-1-\epsilon}\int_0^1 dt_1\,t_1^{-1-\epsilon}\,(1+t_1)^{-1}\;.
\end{eqnarray*}
For more complicated functions, this procedure may have to be iterated, 
but the principle is simple and easily automated.
This is particularly true for multi-loop integrals because they have, 
after Feynman parametrization and integration over the loop momenta, 
the following universal form ($L$ is the number of loops, $N$ the number of 
propagators and $D$ the space-time dimension)
\begin{eqnarray}
G &=& (-1)^N\Gamma(N-LD/2)\int
\limits_{0}^{\infty} \prod\limits_{j=1}^N dx_j\label{loop}\\
&& \delta(1-\sum_{i=1}^N x_i)\,
\frac{{\cal U}(\vec x)^{N-(L+1) D/2}}{{\cal F}(\vec x,\{s,m^2\})^{N-L D/2}}\;,
\nonumber
\end{eqnarray}
where ${\cal U}$ and ${\cal F}$ are polynomials in the Feynman parameters 
and ${\cal F}$ also contains kinematic invariants.
Applying the sector decomposition algorithm\,\cite{Binoth:2000ps} to loop 
integrals in the form (\ref{loop}) isolates the dimensionally regulated 
poles in terms of factorizing Feynman parameters.
Then subtractions of the singularities are carried out, using 
identities like  
\begin{eqnarray}
&&\int_0^1 dx_1\,x_1^{-1+\kappa\epsilon}{\cal F}(x_1,\hat{x})\nonumber\\
&&=\frac{1}{\kappa\epsilon}\int_0^1 dx_1\,{\cal F}(x_1,\hat{x})\,\delta(x_1)+\nonumber\\
&&\int_0^1 dx_1\,x_1^{-1+\kappa\epsilon}\,[{\cal F}(x_1,\hat{x})-{\cal
F}(0,\hat{x})]\;,
\label{plusd}
\end{eqnarray}
where $\hat x=x_2,\ldots,x_N$ and $\lim_{x_1\to 0}{\cal F}(x_1,\hat{x})$
is finite by construction, such that the second term in (\ref{plusd}) is a 
plus distribution:
\begin{eqnarray*}
&&\int_0^1 dx_1\,x_1^{-1+\kappa\epsilon}\,[{\cal F}(x_1,\hat{x})-{\cal
F}(0,\hat{x})]\\
&&=\sum_{n=0}^{\infty}\frac{(\kappa\epsilon)^n}{n!}
\int_0^1 dx_1\,\left[\frac{\ln^n(x_1)}{x_1}\right]_+\,{\cal F}(x_1,\hat{x})\;.
\end{eqnarray*}
Doing these subtractions for all $x_i$ results in a  Laurent series
$${\cal I}=\sum\limits_{k=-2L}^{b}
 { \epsilon}^k\,C_k\;+\;{\cal O}(\epsilon^{b+1})\;,$$
where the order $b$ of expansion in $\epsilon$ is in principle only limited 
by CPU time. 
However, the pole coefficients $C_k$ being sums of complicated 
parameter integrals, their analytical evaluation is in general impossible. 
Therefore they are integrated numerically. For multi-loop integrals 
involving more than one kinematic invariant, Euclidean points have to
be chosen in order to have stable numerics. 
In this way, results have been obtained\,\cite{Binoth:2003ak} for example for 
massless 2-loop 4-point functions with 2 off-shell legs, where no 
analytical results exist yet, 
 all  4-point master integrals needed for the calculation  
of 2-loop Bhabha scattering with massive fermions (analytical results 
exist for two of them\,\cite{Smirnov:2001cm,Smirnov:2004ip,Heinrich:2004iq}), 
two-point-functions with 4 and 5 loops, and for  the planar massless 
3-loop 4-point-function with on-shell legs calculated analytically by 
V.A. Smirnov\,\cite{Smirnov:2003vi}.

\section{PHASE SPACE INTEGRALS}

The phase space integration for the production of 
$N$ massless particles $q\to p_1,\ldots,p_N$ can be written as
\begin{eqnarray}
&&\int d\Phi_{1\to N}=
(2\pi)^{ N - D (N-1)} \nonumber\\
&&\int  \prod\limits_{j=1}^{N} d^Dp_j \,\delta^+(p_j^2) 
\delta\Bigl(q-\sum\limits_{i=1}^{N} p_i \Bigr)\nonumber\\
&&=(2\pi)^{ N - D (N-1)}\, 2^{1-N}\nonumber\\
&&\int  \prod\limits_{j=1}^{N-1} d^{D-1} \vec{p}_j\,\frac{\Theta(E_j)}{E_j}
 \, \delta^+([q-\sum\limits_{i=1}^{N-1} p_i]^2)\;.\nonumber
\end{eqnarray}
At this point one could pick a particular frame and integrate over 
energies $E_j$ and angles $\theta_j$, but for our purposes it is more 
convenient to integrate over the scaled invariants 
$s_{ij}/q^2,\, s_{ij}=(p_i+p_j)^2$, because 
in this way the singularities are located at the origin of parameter space 
and no particular axis is preferred. 
The transformation to the integration variables 
\begin{eqnarray*}
&&x_1=s_{12}/q^2, x_2=s_{13}/q^2, 
x_3=s_{23}/q^2,\\
&&x_4=s_{14}/q^2,x_5=s_{24}/q^2,
x_6=s_{34}/q^2,\,\ldots
\end{eqnarray*}
introduces a Jacobian which is proportional to the square root of the 
determinant of the Gram matrix $G_{ij}=2p_ip_j$. 
The phase space then takes the form\footnote{Note that 
for $N\ge 6$ $\det G$  is zero for 4-dimensional momenta 
because, after elimination of $p_6$ by momentum conservation, 
the vectors $p_1$ to $p_5$ will still be linearly dependent. 
Therefore we only consider the case $N<6$ here.}
\begin{eqnarray}
&&\int d\Phi_{1\to N}=
 C_{\Gamma}^{(N)}(q^2)^{(N-1)D/2-N}\,
\int  \prod\limits_{j=1}^{n_s} dx_j\nonumber\\
&&\delta(1-\sum\limits_{i=1}^{n_s} x_i)
 \Bigl[\Delta_N(\vec{x})\Bigr]^{\frac{D-(N+1)}{2}}
\Theta(\Delta_N)\label{psN}\\
&&\nonumber\\
&&n_s=N(N-1)/2\nonumber\\
&&\Delta_N=|\det G|\,(q^2)^{-N}\nonumber\\
&&C_{\Gamma}^{(N)}=(2\pi)^{N - D(N-1)} 2^{1-ND/2}\nonumber\\
&&\times \,V(D-1)\ldots V(D-N+1)\nonumber\\
&&V(D)=2\pi^{\frac{D}{2}}/\Gamma(\frac{D}{2})\;.\nonumber
\end{eqnarray}

\subsection{$1\to 4$ phase space}
As an example, let us consider the integration of some squared matrix element 
$|M_4|^2$ over the $1\to 4$ partonic phase space, relevant for the 
calculation of $e^+e^-\to 2$\,jets at NNLO:
\begin{eqnarray}
&&\int d\Phi_{1\to 4}=C_{\Gamma}^{(4)}\,(q^2)^{3D/2-5}\nonumber\\
&&
\int\prod\limits_{j=1}^6 dx_j\,\delta(1-\sum\limits_{i=1}^6 x_i)\;|M_4|^2\nonumber\\
&&\left[-\lambda(x_1x_6,x_2x_5,x_3x_4)\right]^{-1/2-\epsilon}
\Theta(-\lambda) \label{4ps}\\
&&\nonumber\\
&&\lambda(x,y,z)=x^2+y^2+z^2- 2\,(xy+xz+yz)\;.\nonumber
\end{eqnarray}
The matrix element is of the form 
\begin{eqnarray*}
|M_4|^2 &\sim&  
\frac{{\cal P}_1(\vec{x},\epsilon)}{(x_2+x_4+x_6)(x_3+x_5+x_6)x_4}\\
&&+\frac{{\cal P}_2(\vec{x},\epsilon)}{x_2(x_2+x_4+x_6)^2}
+\ldots\;,\\
\end{eqnarray*}
where the ${\cal P}_k(\vec{x},\epsilon)$ are some polynomials in the variables $x_i$.
We again see the sums of Feynman parameters in the denominator, 
corresponding to triple invariants $s_{ijk}$, 
giving rise to an overlapping structure. Therefore, the form of the 
integral (\ref{4ps}) is very similar to  the one in eq.\,(\ref{loop}) 
for loop integrals and the overlapping singularities can be disentangled 
by the same principle. However, there are also very important 
differences to loop integrals. The most important one consists 
in the fact that in phase space integrals, 
non-polynomial structures (square roots) appear. 
For example, solving the constraint $-\lambda>0$ in (\ref{4ps}) for $x_6$ 
leads to $x_6^-<x_6<x_6^+$ with 
$x_6^{\pm}=(\sqrt{x_2x_5}-\sqrt{x_3x_4})^2/x_1$.
The substitution $x_6\to (x_6^+-x_6^-)\,y_6+x_6^-$  
remaps the integration range of $x_6$ to an integral from 0 to 1 again
and factorizes the $\lambda$-term:  
\begin{eqnarray*}
&&[-\lambda]^{-1/2-\epsilon}=[x_1^2(x_6^+-x_6)(x_6-x_6^-)]^{-1/2-\epsilon}\\
&&\quad\to [\,y_6(1-y_6)\,]^{-1/2-\epsilon}[x_1(x_6^+-x_6^-)]^{-1-2\epsilon}\;.
\end{eqnarray*}
However, it is possible to eliminate the square roots
by quadratic transformations,  
except in  factors like  
$(1-y_6^2)^{-1/2-\epsilon}$, which do not lead to singularities in $\epsilon$
and therefore are not subject to further sector decomposition. 
This nice feature will be spoiled in the $1\to 5$ case. 

The implementation of sector decomposition for the $1\to 4$ phase space 
served for the calculation of all master phase space integrals which 
are needed for any $1\to 4$ process in massless QCD.
These master integrals have been derived and calculated analytically 
as well as numerically in \cite{Gehrmann-DeRidder:2003bm}.

Moreover, the method also can deal with the full matrix element without 
reduction to master integrals. 
This has been demonstrated in\,\cite{Binoth:2004jv}.
To split the calculation into smaller pieces,
one can write the squared matrix element as a sum over different topologies. 
As the calculation is naturally parallelized by this subdivision 
into topologies, the overall runtime is given by the 
most difficult topology, which took 
about 9 hours for a precision of 0.1\% and less than two hours for 
a precision of 1\% on a Pentium IV 2.2 GHz PC.

\subsection{$1\to 5$ phase space}
The $1\to 5$ partonic phase space, relevant for the 
calculation of $e^+e^-\to 3$\,jets at NNLO, involves the 
integration over 9 independent invariants: 
\begin{eqnarray}
&&\int d\Phi_{1\to 5}=C_{\Gamma}^{(5)}
\int\prod\limits_{j=1}^{10} dx_j\label{ps5}\\
&&\delta(1-\sum\limits_{i=1}^{10} x_i)\,
\left[\Delta_5(\vec{x})\right]^{(D-6)/2}\Theta(\Delta_5)\;.\nonumber
\end{eqnarray}
Note that $C_{\Gamma}^{(5)}\sim V(D-4)=
2\pi^{-\epsilon}/\Gamma(-\epsilon)$ is of order $\epsilon$, 
therefore the integral (\ref{ps5}) contains a fake singularity in 
$\left[\Delta_5(\vec{x})\right]^{(D-6)/2}=
\left[\Delta_5(\vec{x})\right]^{-1-\epsilon}$, but this presents no
problem for sector decomposition as the algorithm will extract the singular
factor and the $\epsilon$-expansion subroutine 
will take the prefactor of order $\epsilon$ into account, 
such that the fake singularity will be eliminated automatically. 
What is more of a problem are the non-polynomial structures  
which occur here, because denominators of the form $g(x,y)=a+x+y-\sqrt{a^2+x+y}$, 
where $a$ is a constant, can produce a singularity for $x,y\to 0$ without 
having the right scaling behaviour amenable to sector decomposition. 
The task is to transform such terms away without increasing the complexity
of the integrand too much. It should be noted that the size of the  
expressions in the  $1\to 5$ case is considerably larger than in the $1\to 4$
case, such that it becomes much more important to produce as few subsectors as
possible. 

The simplest example to calculate is the 5-particle phase space volume
without any matrix element. In \cite{Gehrmann-DeRidder:2003bm} 
a general analytic expression for the $1\to N$ phase space volume is given, 
such that the numerical result can be easily checked. 
By sector decomposition, one obtains
\begin{eqnarray*}
\int d\Phi_{1\to 5}
&=&\frac{(4\pi)^{4\epsilon-7}}{\Gamma(1-2\epsilon)\Gamma(2-2\epsilon)}
\Big[0.00347\\
&&+0.05469\epsilon+0.44336\epsilon^2\\
&&+2.47424\epsilon^3+10.7283\epsilon^4 +{\cal O}(\epsilon^5) \Big]
\end{eqnarray*}
which agrees with the analytical result to an 
accuracy of 0.5\% after a runtime of about 10 minutes.

\subsection{One loop plus single real emission}


Apart from the double real emission and the 
two-loop virtual contributions to the cross section of 
$e^+e^-\to jets$ at NNLO, there is also a contribution where 
one-loop virtual corrections are combined with single real emission. 
In this class, the most complicated diagram which can occur 
in the calculation of  $e^+e^-\to 2$\,jets is a
box graph with one off-shell leg.
This type of diagram can easily be calculated by sector decomposition: 
The one-loop box can be expressed by Hypergeometric functions 
$_2F_1(1,-\epsilon,1-\epsilon;x_i/x_j)$.
Then the parameter representation of the Hypergeometric functions
can be used and the resulting one-dimensional parameter integrals 
can be combined with the ones for the  3-particle phase space 
to end up with a 4-dimensional parameter integral 
which can be directly fed into the  sector decomposition routine.

For $e^+e^-\to 3$\,jets, the most complicated one-loop diagrams are 
pentagons with one off-shell leg. These could be reduced to boxes 
by standard reduction techniques\,\cite{bdk,Binoth:1999sp}, 
but as the reduction
introduces inverse determinants of kinematic matrices which may 
lead to numerical instabilities, it is more convenient to 
apply the sector decomposition routine for loop integrals 
directly to the pentagon which is of the form
\begin{eqnarray*}
I_5&=&-\Gamma(3+\epsilon)\int \prod_{i=1}^{5}
dz_i \,\delta(1-\sum_{i=1}^5 z_i)\,{\cal F}^{3+\epsilon}\\
-{\cal F}&=&s_{12}\,z_1z_5+s_{23}\,z_1(z_{3}+z_{4}+z_{5})\\
&+&s_{13}\,z_5(z_{1}+z_{2})+s_{14}\,z_5(z_{1}+z_{2}+z_{3})\\
&+&s_{24}\,z_1(z_{4}+z_{5})+
s_{34}\,(z_{1}+z_{2})(z_{4}+z_{5})\,.
\end{eqnarray*}
After sector decomposition in the variables $z_i$, 
one obtains an expression where the poles of 
the virtual integral already have been extracted: 
\begin{eqnarray*}
I_5&=&\sum_{\alpha=0}^2P_{\alpha}/\epsilon^{\alpha}\;,\\
P_{\alpha}&=&\int_0^1\prod_{i=1}^{4-\alpha}dt_i\, {\cal
G}(t_i,s_{12},\ldots,s_{34})\;,\;\\
&&\lim_{t_i\to 0}{\cal G}\not=0\;.
\end{eqnarray*}
This expression can then be inserted into the 4-particle phase space
and one can  proceed with decomposition in the scaled invariants
$x_1,\ldots,x_6$. 
Note that no problems with thresholds will occur here as the kinematics is such
that all  invariants $s_{ij}$ are non-negative.

\section{SUMMARY AND OUTLOOK}
The automated sector decomposition algorithm is a 
powerful method to isolate overlapping infrared poles and 
to calculate  numerically not only multi-loop integrals, 
but also phase space integrals where 
some of the particles can become theoretically unresolved, leading to 
infrared singularities. 
In particular, the method allows the calculation of 
the one-loop plus single real emission and the double real emission
contribution to $e^+e^-\to 2$ or 3 jets at NNLO
without having to establish a subtraction 
scheme and to integrate analytically 
over complicated subtraction terms. 
The inclusion of a measurement function also does not present a problem, 
as has been demonstrated already in\,\cite{Anastasiou:2004qd},  
such that a fully differential Monte Carlo program can be constructed
based on this method. 
The only drawback of the method is the fact that it
generates a large number of functions, but  it has been shown already
that in the case of  $e^+e^-\to 2$ jets at NNLO, this does not lead to 
unacceptable integration times. Further, the functions are 
numerically well-behaved by construction. 
How the NNLO calculation of the process $e^+e^-\to 3$ jets with this method 
performs numerically will turn out in the near future.

The generalization to other processes than $e^+e^-$ annihilation is feasible,
but cases where some of the kinematic invariants take negative values 
cannot be treated without further development of the method.

\section*{Acknowledgements}

I would like to thank the organizers of Loops and Legs 2004 
for the invitation 
and for the interesting conference.
This work  was supported 
by the Bundesministerium f\"ur Bildung und Forschung
through Grant No.\ 05~HT4GUA/4.

\end{document}